\documentclass[unsortedaddress,preprint,aps,show pacs,showkeys]{revtex4}
\usepackage{graphicx}
\usepackage{dcolumn}
\usepackage{amsmath}
\usepackage{amssymb}
\usepackage{amsfonts}
\usepackage{bm}
\usepackage{color}
\newcommand{\be}{\begin{equation}}
\newcommand{\ee}{\end{equation}}
\newcommand{\ba}{\begin{eqnarray}}
\newcommand{\ea}{\end{eqnarray}}
\begin{document}
\title{Phase behavior of a double-Gaussian fluid\\displaying water-like features}
\author{Cristina Speranza\footnote{Email: {\tt csperanza@unime.it}}, Santi Prestipino\footnote{Corresponding author. Email: {\tt sprestipino@unime.it}}, Gianpietro Malescio\footnote{Email: {\tt malescio@unime.it}}, and Paolo V. Giaquinta\footnote{Email: {\tt paolo.giaquinta@unime.it}}}
\affiliation{Universit\`a degli Studi di Messina, Dipartimento di Fisica e di Scienze della Terra, Contrada Papardo, I-98166 Messina, Italy}
\date{\today}
\begin{abstract}
Pair potentials that are bounded at the origin provide an accurate description of the effective interaction for many systems of dissolved soft macromolecules (e.g., flexible dendrimers). Using numerical free-energy calculations, we reconstruct the equilibrium phase diagram of a system of particles interacting through a potential that brings together a Gaussian repulsion with a much weaker Gaussian attraction, close to the thermodynamic stability threshold. Compared to the purely-repulsive model, only the reentrant branch of the melting line survives, since for lower densities solidification is overridden by liquid-vapor separation. As a result, the phase diagram of the system recalls that of water up to moderate (i.e., a few tens MPa) pressures. Upon superimposing a suitable hard core on the double-Gaussian potential, a further transition to a more compact solid phase is induced at high pressure, which might be regarded as the analog of the ice I to ice III transition in water.
\end{abstract}
\pacs{61.20.Ja, 64.70.dm, 64.60.Q-}
\keywords{double-Gaussian potential; water phase diagram; water-like anomalies}
\maketitle

\section{Introduction}

One of the central problems in the statistical mechanics of classical systems is understanding how the phase behavior of particles is affected by the details of their mutual interactions. Up until about a couple of decades ago, this appeared as an essentially solved problem for at least monatomic substances characterized by isotropic interactions. For such systems the typical phase diagram was considered to be argon-like, featuring a liquid-vapor critical point, a solid-liquid-vapor triple point, and a melting temperature monotonously increasing with pressure. These features are strictly related to the essential ingredients of the interparticle forces in atomic systems, namely: (i) a harsh short-range repulsion caused by the overlap of the outer electronic shells and (ii) an attraction at larger separations arising from multipole dispersion forces, as embodied, e.g., in the popular Lennard-Jones potential.

Starting from the early 1990Õs, the interest for new materials such as fullerenes and soft matter (colloids, polymers, dendrimers, surfactants, etc.) has led to explore the phase behavior of model potentials that do not fit the Lennard-Jones paradigm. Pair potentials devoid of any physical meaning in the context of microscopic interactions of atomic systems would acquire a physical relevance when assumed to describe {\em effective} interactions among macromolecules. In fact, due to their molecular architecture, complex-fluid particles have internal degrees of freedom which can be accounted for in a coarse-grained fashion by an effective center-of-mass potential~\cite{Likos,Likos4}. It was found that deviations of the potential from the typical argon-like form may yield surprising novelties in the system phase diagram, such as: (i) the disappearance of the liquid phase for a hard-core repulsion augmented with a sufficiently short-ranged attractive tail~\cite{Bolhuis}, (ii) the existence of two distinct liquid phases with a liquid-liquid critical point for ``softened'' interactions where a hard-core potential is supplemented by a finite repulsive shoulder (see e.g. Ref.\,\cite{Franzese,Wilding,Gibson}), and (iii) either a point of maximum melting temperature followed by a line of reentrant melting or no upper bound in the melting temperature and stable cluster crystals at high pressure for bounded repulsions~\cite{Likos2,Prestipino1,Zhang}.

Here we investigate the behavior of a model system of particles interacting through a bounded repulsion plus a weak attractive tail, which is thought to be somehow representative of the effective interaction between polymer coils in a not too concentrated solution~\cite{Louis,Likos3}. We consider the double-Gaussian model (DGM), in which both the repulsive and attractive components of the potential are chosen of Gaussian shape. Moreover, the attractive well is shifted to outside the repulsive core so that the strength of the repulsion is roughly unaffected by the attraction. Due to the finiteness of the repulsion, when the attraction overcomes a certain strength the system becomes thermodynamically unstable. Conditions for thermodynamic stability were first derived by Fisher and Ruelle~\cite{Ruelle,Fisher} and then implemented by Heyes and Rickayzen~\cite{Heyes}.

In a previous paper~\cite{Prestipino2}, we studied the equilibrium behavior of a DGM system with an extremely weak attraction, which could essentially be viewed as a small perturbation of the repulsive interaction. This nonetheless, the phase diagram turned out to be considerably more complex than for the Gaussian-core model (GCM) (i.e., no attraction whatsoever)~\cite{Prestipino1}, with four solid phases, two fluid phases, and two distinct reentrant-melting lines. We hereby consider a four-times stronger attraction, near the stability threshold of the DGM. This system provides an example of an as yet unexplored regime in which a bounded repulsion coexists with a fairly strong attraction. It would especially be interesting to see how effective the finite repulsion can be with regard to condensation and crystallization. Given the peculiar nature of the interparticle forces, a number of questions arise concerning the location and extension of the liquid-vapor (phase coexistence) region, as well as the topology of the melting line and of the loci of water-like anomalies. Anticipating our results, we find that a strong enough Gaussian attraction causes the liquid-vapor region to widen at the expenses of the (bcc) solid phase, which in the present case is found to always melt into a denser liquid. The resulting scenario is vaguely reminiscent of water phase behavior in the low-to-moderate pressure regime. To strengthen the resemblance with water, a suitable hard core can be added to the DGM potential in order to induce the transition to a more compact solid phase at high pressure, somewhat akin to the transformation occurring in water from ice I to ice III at about 200 MPa, while the low-pressure behavior is left unchanged.

The outline of the paper is as follows. After recalling the model definition and briefly describing the simulation method in Section II, we present and discuss our results in Section III. Some concluding remarks are postponed to Section IV.

\section{Model and method}
\setcounter{equation}{0}
\renewcommand{\theequation}{2.\arabic{equation}}

The general double-Gaussian potential has been defined in \cite{Prestipino2} as\be
u(r)=\epsilon\exp\{-r^2/\sigma^2\}-\epsilon_2\exp\{-(r-\xi)^2/\sigma_2^2\}\,,
\label{2-1}
\ee
where $r$ denotes the interparticle distance and $\epsilon$ and $\sigma$ are arbitrary energy and length units, respectively. In Ref.~\cite{Prestipino2}, we considered a slightly perturbed GCM fluid by taking $\epsilon_2=0.005\,\epsilon,\sigma_2=\sigma$, and $\xi=3\sigma$ in Eq.\,(\ref{2-1}). With this choice, the repulsive core is barely affected by the attraction and the high-pressure behaviors of the GCM and DGM systems are similar. Indeed, save for reduced pressures lower than 0.0005, the DGM melting line runs very close to that of the GCM system, reaching a maximum value of about $0.010\,\epsilon/k_B$ at about the same density. Moreover, two distinct fluid phases are found, liquid and vapor, which can coexist up to a critical temperature of $0.0077\,\epsilon/k_B$.

As recalled in the Introduction, a problem of stability may arise for a system of particles interacting through a bounded repulsive potential endowed with an attractive tail. If the attraction is too slowly decaying or too strong, in the infinite-size limit the system collapses in a {\em finite} region of space and no clearly-defined thermodynamics is possible. To figure out whether this be the case or not for the DGM potential, we resorted to a pair of criteria originally put forward by Ruelle~\cite{Ruelle} (quoted as Theorems 1 and 2 in Ref.~\cite{Heyes}). If $\widetilde{u}(k)$ denotes the Fourier transform of the potential, a sufficient condition for instability is $\widetilde{u}(0)<0$. On the other hand, the positivity of $\widetilde{u}(k)$ is enough to conclude that the system is stable. We showed in \cite{Prestipino2} that, for $\sigma_2=\sigma$ and $\xi=3\sigma$, these conditions jointly lead to a stable DGM system as long as $\epsilon_2<0.026315\,\epsilon$. For the present study, we have chosen $\epsilon_2=0.02\,\epsilon$, i.e., four times larger than in the previously investigated case; this would now make the critical temperature $T_c$ definitely larger than the maximum melting temperature.

In order to perform the simulation study, we first need to identify the relevant crystal structures. This is accomplished through exact $T=0$ total-energy calculations for the same candidate structures as considered in \cite{Prestipino2}. We find that the body-centered-cubic (bcc) crystal provides the lowest chemical potential among all structures for all values of the pressure $P$, any other structure being much higher in chemical potential to be deemed relevant for non-zero temperatures.

The phase diagram of the DGM system was carefully investigated by Monte Carlo (MC) simulation in the $NPT$ (isothermal-isobaric) ensemble, with $N=1024$ particles and periodic boundary conditions. In order to draw the liquid-vapor line, we carried out Gibbs-ensemble simulations~\cite{Panagiotopoulos} of samples ranging from 1000 to 4000 particles, depending on the temperature. Typically, as many as $10^5$ sweeps were generated at equilibrium for each $(T,P)$ state, which turned out to be sufficient to obtain accurate statistical averages of the volume and energy per particle. Much longer runs of $5\times 10^5$ sweeps each were performed in order to compute the chemical potential of the fluid phase by Widom's particle-insertion method~\cite{Widom}. The location of the melting transition was determined through thermodynamic integration of chemical-potential derivatives along isobaric and isothermal paths (see e.g. Ref.~\cite{Saija1}), connecting the system of interest to a reference system whose free energy is already known. While the reference state for the fluid phase was a dilute gas, in the solid region we took a low-temperature bcc crystal as the starting point of an MC trajectory. In any such state, the Helmholtz free energy per particle was computed by the Einstein-crystal method~\cite{Frenkel,Polson}.

\section{Results}
\setcounter{equation}{0}
\renewcommand{\theequation}{3.\arabic{equation}}

We show in Fig.\,1 the $T$-$P$ phase diagram of the DGM system with $\epsilon_2=0.02\epsilon,\sigma_2=\sigma$, and $\xi=3\sigma$, as derived from our numerical free-energy calculations (in order to improve the visibility of the low-pressure region, we have unevenly stretched the scale on the pressure axis by reporting the square root of $P$ rather than $P$ itself). The same DGM phase portrait but in the $\rho$-$T$ plane is displayed in Fig.\,2, so as to highlight the coexistence regions. Finally, we zoom on the triple-point region in Fig.\,3, which also shows a comparison between the ``exact'' transition lines obtained from simulation and the approximate ones derived by a number of faster theoretical approaches (see below).

Compared to the case investigated in Ref.\,\cite{Prestipino2}, the critical temperature of the present DGM system is much higher ($0.0657$ vs. $0.0077$ in reduced, $\epsilon/k_B$ units), due to a stronger attraction. The critical values of the number density and pressure are estimated to be $\rho_c=0.0525$ and $P_c=0.00102$ (from now on, all quantities are given in reduced units). The extent of the liquid-vapor region can be appreciated in Fig.\,2; interestingly, the region of liquid-vapor coexistence is well correlated with the set of $(T,P)$ points where the homogeneous hypernetted-chain (HNC) equation could not be solved~\cite{Hansen}.

At variance with the equilibrium between liquid and vapor, which is essentially dictated by the value of $\epsilon_2$, the location of the melting line is less sensitive to the change in $\epsilon_2$ from 0.005 to 0.02, at least as long as $\epsilon_2\ll\epsilon$. Hence, the widening of the liquid-vapor region with increasing $\epsilon_2$ goes at the expenses of the regular-melting branch, which becomes accordingly shorter. This is confirmed in Figs.\,1-3, where we see that the ``regular'' branch of the melting line is even absent in the present case. Thus, the bcc crystal always melts into a denser liquid at all temperatures below the triple-point temperature ($T_t=0.01087$). In Fig.~3, we show the melting point of a metastable bcc crystal of negative pressure as well as data for the bcc density along the sublimation line, computed through $P=0$ simulations in the crystal (the actual bcc-vapor coexistence pressures are indeed smaller than $10^{-6}$). At $T_t$, the $P=0$ bcc density is only slightly smaller than the liquid density, which complies with the indication coming from our free-energy data. In Fig.\,3, we have also included the melting and freezing points obtained, at various temperatures, through the variational theory introduced in Ref.\,\cite{Prestipino2}, as well as a number of bcc melting points estimated by the so-called heat-until-it-melts (HUIM) method, also described in \cite{Prestipino2} (the degree of solid overheating is 13\% near the triple point but much larger -- about 23\% -- at the highest probed pressures). Qualitatively speaking, the agreement of both approximations with the MC points is fairly good.

The phase behavior illustrated so far calls to mind water near its triple point. As discussed many times in the literature, in liquid water the hydrogen bond brings forth two (open and compact) local structures in competition with each other. At a coarse-grained level, this interplay may be described through a two-body soft-core repulsion, which may then be called responsible for the anomalous slope of the melting line, whereas the behavior for larger pressures (say, above one hundred MPa) would only probe shorter-length features in the effective pair potential. In the last few decades, isotropic pair potentials with two repulsive scales were the leading paradigm for understanding water at a simplified level, especially with regard to its many thermodynamic, structural, and dynamic anomalies~\cite{Hemmer,Young,Debenedetti,Jagla,Sadr-Lahijani,Franzese,Wilding}. However, it is worth recalling that anomalies are also observed in the phase behavior of particles interacting through a one-scale bounded potential like the GCM potential~\cite{Mausbach,Krekelberg,Speranza}.

For a better quantitative comparison of the present DGM system with water, we note that the DGM value of $T_c/T_t$ is 6.05, to be contrasted with the water value of 2.37 (since it is difficult to extract $P_t$ from our data, we have no reliable estimate of $P_c/P_t$ for the DGM system; however, its order of magnitude is the same as for water). On a more physical ground, it would be crucial to ascertain whether the morphology of water-like anomalies in the DGM liquid is similar to that of water. To this end, we have monitored the maximum of the density $\rho$ as well as the minimum of the isobaric specific heat $c_P$ as a function of pressure (see Fig.\,1). In the DGM liquid, the volumetric anomaly occurs beyond a certain pressure only, at distinct variance with liquid water where a maximum-density point is found only {\em below} a certain pressure~\cite{Chaplin}. We safely exclude that, upon lowering the pressure below  $P=0.03$, the maximum-density line eventually bends towards higher pressures, thus revealing the existence of a density-minimum locus deep inside the solid region~\cite{Poole,Speranza} (in cooling the liquid, we kept on recording density data until the solid nucleated spontaneously; therefore, the maximum-density line terminates just where it intersects the liquid spinodal line). The continuation of the maximum-density locus on the high-pressure side would likely go similarly as in the GCM, where the temperature of the maximum-density point attains a maximum slightly above $P=1$ and then decreases for higher pressures~\cite{Mausbach}. The specific-heat minimum of the DGM liquid shifts to higher temperatures as the pressure increases, again contrary to water where the specific-heat minimum moves to lower temperatures upon compression~\cite{Wagner}. Next, we computed the so-called pair entropy $s_2$ (see e.g. \cite{Saija1}) along a number of isotherms, and the pressure points at which $-s_2$ attains its maximum are reported in Fig.\,1. Like for many softly-repulsive potentials, the locus of the $-s_2$ maximum (``structural anomaly'') is found to originate near the point of maximum melting temperature (the triple point here), pointing towards higher temperatures upon compression (the finiteness of the potential here forbids the further possibility of a $-s_2$ minimum).
Finally, we investigated the anomalous behavior of the self-diffusion coefficient $D$, as computed from the mean square displacement of a particle in a molecular-dynamics run. In a normal fluid $D$ is expected to steadily decrease upon isothermal compression, whereas in an anomalous liquid a clearcut minimum (followed by a maximum for higher pressures) is found. However, for particles interacting via a bounded potential a maximum value of $D$ is never reached and only a minimum is detected. The $(T,P)$ locus of the minimum-$D$ points is plotted in Fig.\,1. Like in the GCM~\cite{Krekelberg}, the line of diffusion anomaly lies between those of structural and density anomalies.

We have seized the opportunity offered by a particle system with a precisely known phase diagram to test the efficacy of a popular criterion of ordering, based on the calculation of the residual multiparticle entropy (RMPE) (see e.g. \cite{Giaquinta1} and references therein). The RMPE weighs up the contribution of multi-particle (threefold and more) spatial correlations to the configurational entropy of the fluid -- see for instance Refs.\,\cite{Prestipino3,Prestipino4}. As originally discussed in Ref.~\cite{Giaquinta2} in the context of hard spheres and later observed for other interactions too, the zero-RMPE condition provides a useful method to unveil hidden tendencies to ordering in a fluid phase (i.e., a positive RMPE would be strong evidence of a highly-structured liquid prone to freezing). The density evolution $\Delta s(\rho)$ of the RMPE within the liquid phase is plotted in Fig.\,4 for various temperatures $T$, ranging from slightly below $T_t$ to $T_c$ and beyond. There are two separate ranges of densities where $\Delta s$ approaches zero value from below: one connected with the condensation transition and another, at lower temperature, instead related to freezing. For $T>T_c$, the RMPE is always negative but $\Delta s(\rho)$ shows a sharp maximum, only slightly negative, roughly on top of the continuation of the liquid-vapor coexistence line (i.e., on the Widom line). For lower densities, $\Delta s$ becomes increasingly negative on entering more and more deeply into the vapor phase, as a result of the increasing loss of correlations between the particles. Below $T_c$, $\Delta s(\rho)$ is slowly varying inside the liquid phase, remaining negative as the density approaches the boiling-point value and probably also in the whole metastable-liquid region. This is in marked contrast with the expectation, based on integral-equation theories, of Ref.\,\cite{Giaquinta3} that the RMPE blows up to $+\infty$ when the liquid-vapor spinodal line is approached from the liquid side. Slightly below $T_t$, the RMPE eventually vanishes upon decompression, along a line that lies not far from the freezing line (like the freezing temperature, also the zero-RMPE temperature is a monotonously-decreasing function of the pressure). Similarly to the GCM case~\cite{Giaquinta1}, the locus of points where the RMPE vanishes lies entirely within the solid region.

To expand the contact between the present model system and water far beyond the triple-point region, the DGM potential should be deeply modified in the inner core. For example, if a suitably short-ranged hard core is added to the DGM potential, a phase transition to a more compact crystal phase will eventually be induced at higher pressure, with reasonably little influence on the low-pressure characteristics. Since a proof of concept is enough here, we decided to locate the liquid-vapor binodal line by the simple HNC equation (roughly as the temperature locus within which no solution is found to the HNC equation) while employing the simple criterion introduced in Refs.\,\cite{Prestipino5,Prestipino6} to draw the melting line $T_m(\rho)$. The latter criterion combines the standard Lindemann rule with a description of the solid phase as an elastic continuum; the alleged melting locus is the upper envelope of the melting lines drawn for each single crystal lattice. Past experience has shown that the melting line predicted by this criterion is qualitatively accurate, even though the $T_m$ values are typically overestimated by a factor of two. With this procedure, we obtain the schematic phase diagrams shown in Fig.\,5. They refer to the present DGM fluid (left panel) and to a modified DGM (mDGM) system where the DGM potential has been augmented with an inverse power of the distance (right panel) -- for our demonstration, we took a diverging repulsion of the form $\epsilon(\sigma'/r)^{24}$ with $\sigma'=\sigma$. We see that the change in the potential only affects the high-density behavior (say, $\rho>0.3$); moreover, the modification is precisely that expected, with the bcc phase being eventually overcome in stability by both the hcp and fcc phases, the latter phase being the most stable for high densities (as it is already for $T=0$). The overall phase portrait is now more closely reminiscent of water, with the bcc-fcc coexistence line nicely mimicking the locus of points separating ice I from ice III (the bcc-fcc line moves to higher pressures upon lowering $\sigma'/\sigma$). Clearly, the resemblance of the mDGM fluid to water is only superficial, lying exclusively in the similar appearance of the phase-transition lines. This similarity is not even perfect, considering the apparent non-monotonicity of the melting line of the low-pressure mDGM solid.

\section{Conclusions}

In this paper, we have analyzed the phase behavior of a system of particles interacting via a double-Gaussian model (DGM) potential with a fairly strong attraction, close to the thermodynamic stability threshold. Compared to a similar system with a more modest attraction strength~\cite{Prestipino2}, there is a unique solid (bcc crystal) phase which always melts into a slightly denser liquid, with a much higher $T_c/T_t$ ratio of about 6. Hence, there is a whole range of temperatures where, under isothermal compression, the (vapor) system is led to condensate but not to eventually become solid. The ensuing scenario is reminiscent of water behavior at low pressure, though water-like anomalies do not occur in similar terms in the two systems. In particular, the slope of the maximum-density locus of the DGM liquid has the wrong sign compared to water. In order to build up an isotropic fluid with more resemblance to water, we have supplemented the DGM potential with a further length scale, by adding a suitably chosen hard core, showing that this generates a further phase transition at higher pressure to a more compact fcc crystal.

\newpage
\section*{Figure captions}
%
%
{\bf FIGURE 1}. (Color online). Numerical phase diagram of the DGM system in the $T$-$P$ plane. We show: the liquid-vapor coexistence points obtained from Gibbs-ensemble simulations (high-$T$ black dots), along with the estimated critical point (black asterisk); the solid-liquid coexistence points obtained from the ``exact'' free-energy calculations described in the text (low-$T$ black dots); the structural-anomaly locus (red open diamonds), the volumetric-anomaly locus (red open triangles), the diffusion-anomaly locus (red open squares), and the minimum-$c_P$ locus (red open inverted triangles). The lines through the points are plotted as a guide for the eye. On the scale of the figure, the bcc-vapor coexistence line is invisible.

%
%
{\bf FIGURE 2}. (Color online). Numerical phase diagram of the DGM system in the $\rho$-$T$ plane. We show: a number of points along the liquid-vapor binodal line obtained by Gibbs-ensemble simulations (black dots), along with the estimated critical point (black asterisk); a number of points along the line (dubbed ``HNC pseudospinodal'' in the figure) enclosing the region where the HNC equation could not be solved (blue open dots); the solid-liquid coexistence points obtained from the ``exact'' free-energy calculations described in the text (black diamonds and squares); some points on the $P=0$ isobar of the bcc solid (black right-pointing triangles). The lines through the points are plotted as a guide for the eye. Note that the low-density region of the plot (i.e., the one enclosed in the red frame) is shown magnified in the next Fig.\,3.

%
%
{\bf FIGURE 3}. (Color online). DGM phase diagram in the $\rho$-$T$ plane: Magnification of the low-density region. We show: two points on the liquid-vapor binodal line obtained by Gibbs-ensemble simulations (black dots); two points along the line enclosing the region where the HNC equation could not be solved (blue open dots); a few solid-liquid coexistence points obtained from the ``exact'' free-energy calculations described in the text (black diamonds and squares); some points on the bcc sublimation line, obtained from MC simulations of the bcc crystal at $P=0$ (black right-pointing triangles); the HUIM melting points obtained by heating (in steps of $\Delta T=0.0002$ close to melting) an originally perfect bcc crystal along isochoric and isobaric paths until it melted (blue plusses and crosses, respectively); the variational-theory (VT) coexistence points (blue open diamonds and squares); the locus of zero-$\Delta s$ points (blue open triangles). The lines through the points are plotted as a guide for the eye. The yellow-shaded regions denote two-phase coexistence regions. In the inset, a zoom on the triple-point region is provided.

%
%
{\bf FIGURE 4}. (Color online). DGM-system RMPE, i.e., $\Delta s=s^{ex}-s_2$, defined as the difference between the excess and two-body entropies per unit particle, plotted as a function of the density along a number of isotherms (see legend).

%
%
{\bf FIGURE 5}. (Color online). Schematic phase diagram of the DGM fluid with $\epsilon_2=0.02\epsilon$ (left) and of the modified DGM (mDGM) system defined in the text (right). These diagrams were obtained using the HNC approximation in combination with a Lindemann-type criterion of melting\,\cite{Prestipino3}. The region delimited from above by the black dots is the $(\rho,T)$ set of points where the HNC equation has no solution. It turns out that this region is almost insensitive to the inner part of the core. The lines are tentative melting loci for the fcc (blue dotted line), the bcc (red solid line), and the hcp crystal (cyan dashed line). More realistically, the actual values of $T_m$ are roughly a half of those shown. The inclusion of an inner ``hard'' core has led to the stabilization of the fcc crystal at high pressure.

\newpage
%
%
\begin{figure}
\centering
\includegraphics[width=18cm]{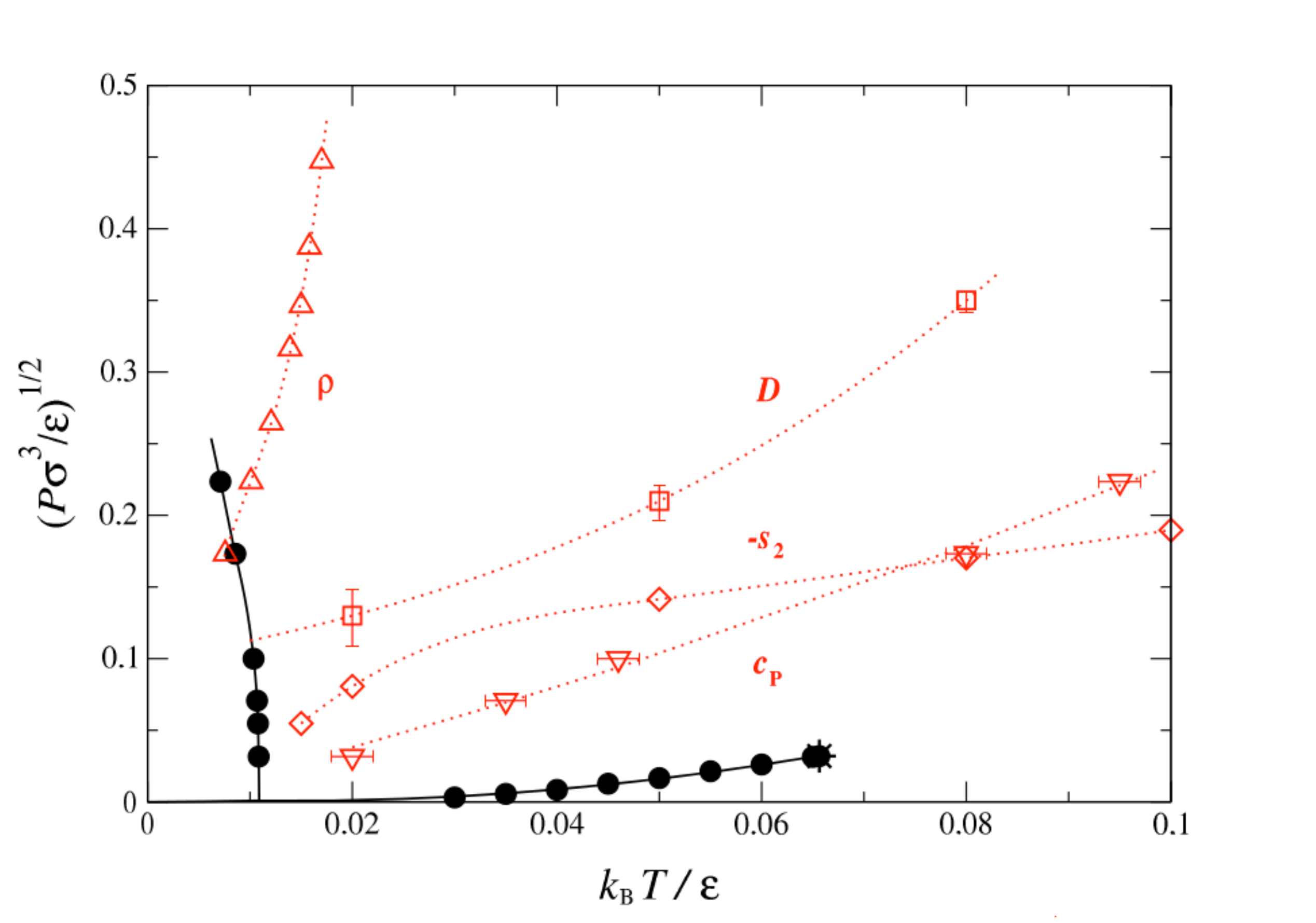}
\caption{
}
\label{fig1}
\end{figure}

%
%
\begin{figure}
\centering
\includegraphics[width=18cm]{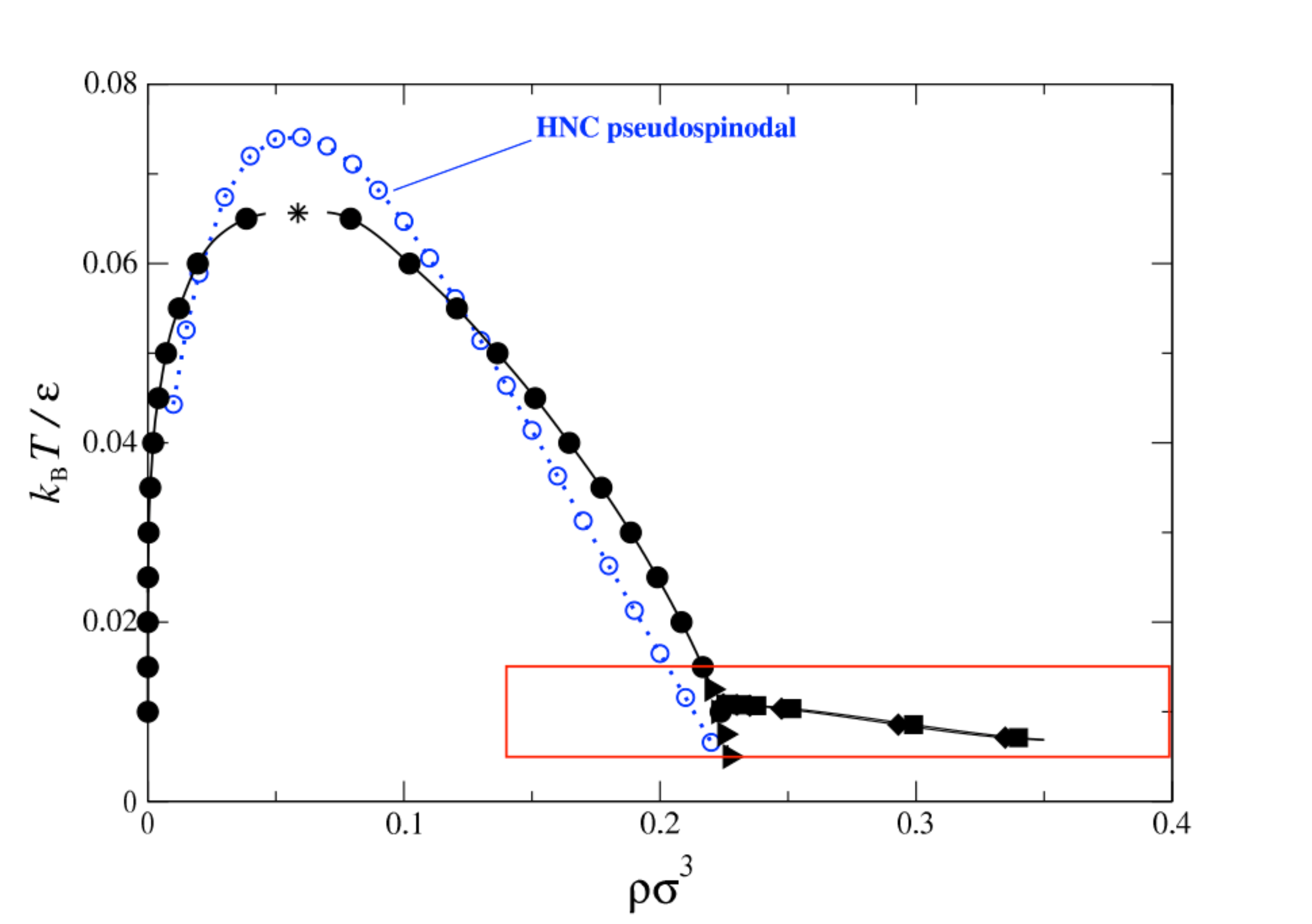}
\caption{
}
\label{fig2}
\end{figure}

%
%
\begin{figure}
\centering
\includegraphics[width=18cm]{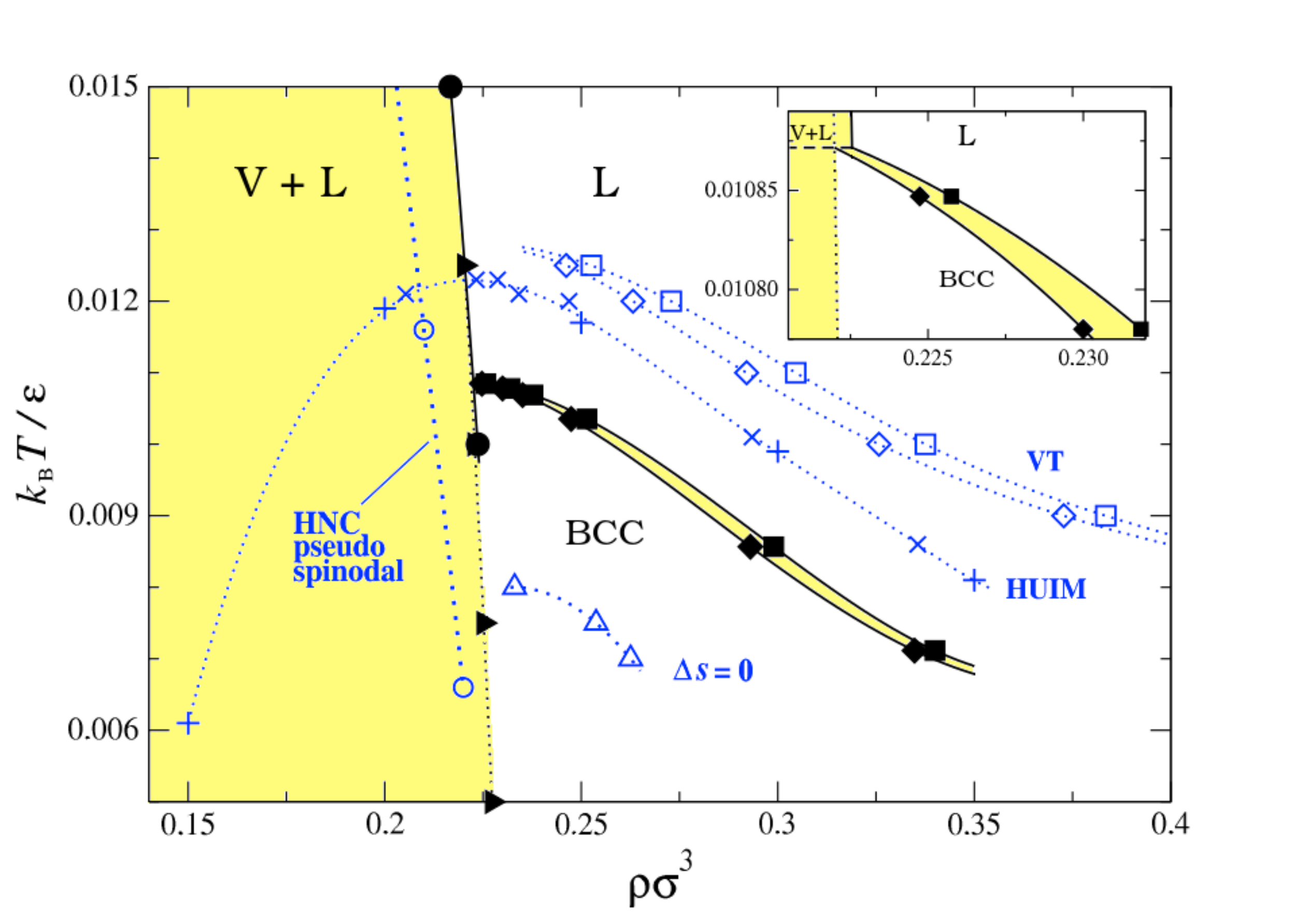}
\caption{
}
\label{fig3}
\end{figure}

%
%
\begin{figure}
\centering
\includegraphics[width=18cm]{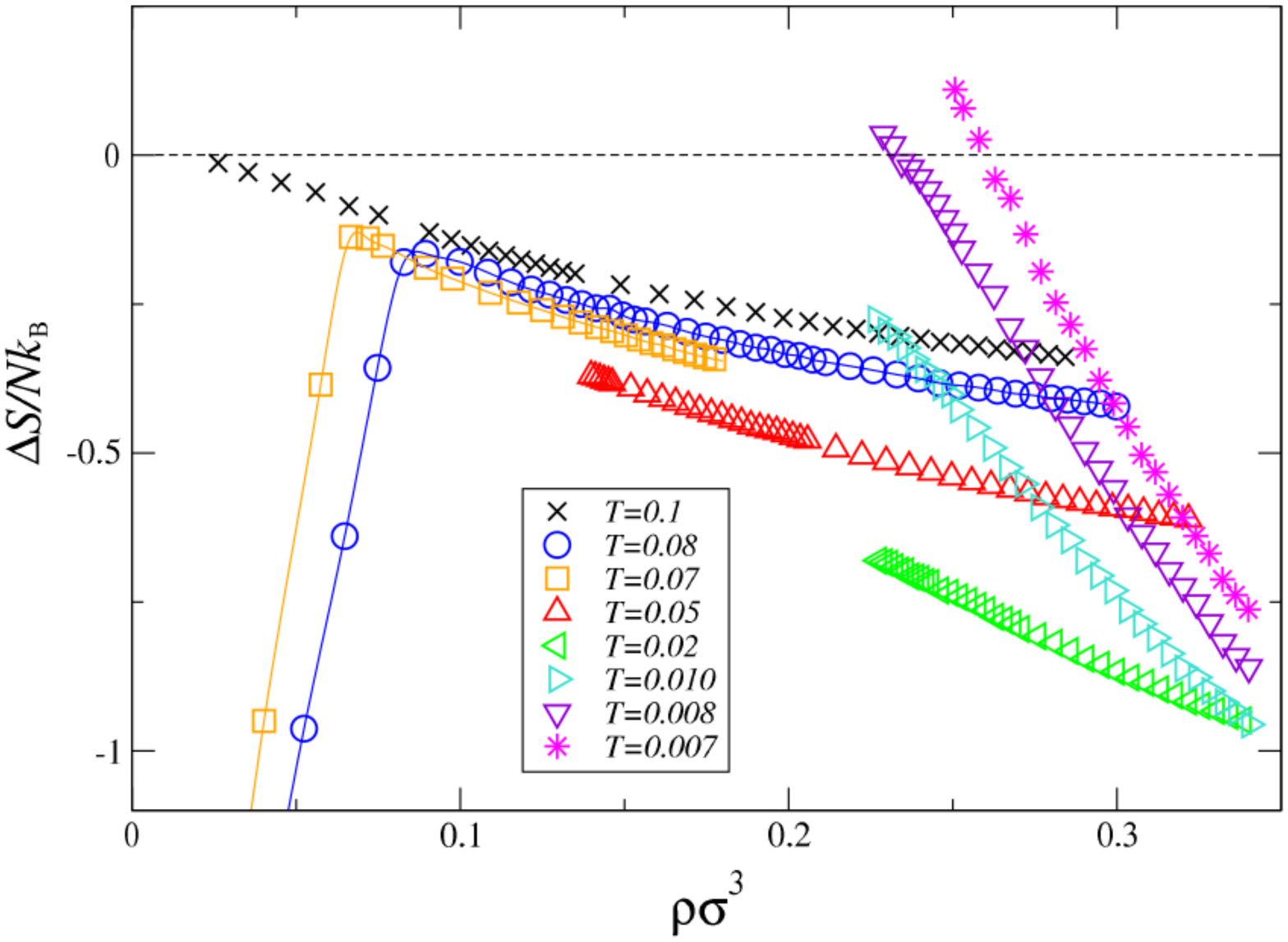}
\caption{
}
\label{fig4}
\end{figure}

%
%
\begin{figure}
\centering
\includegraphics[width=16cm]{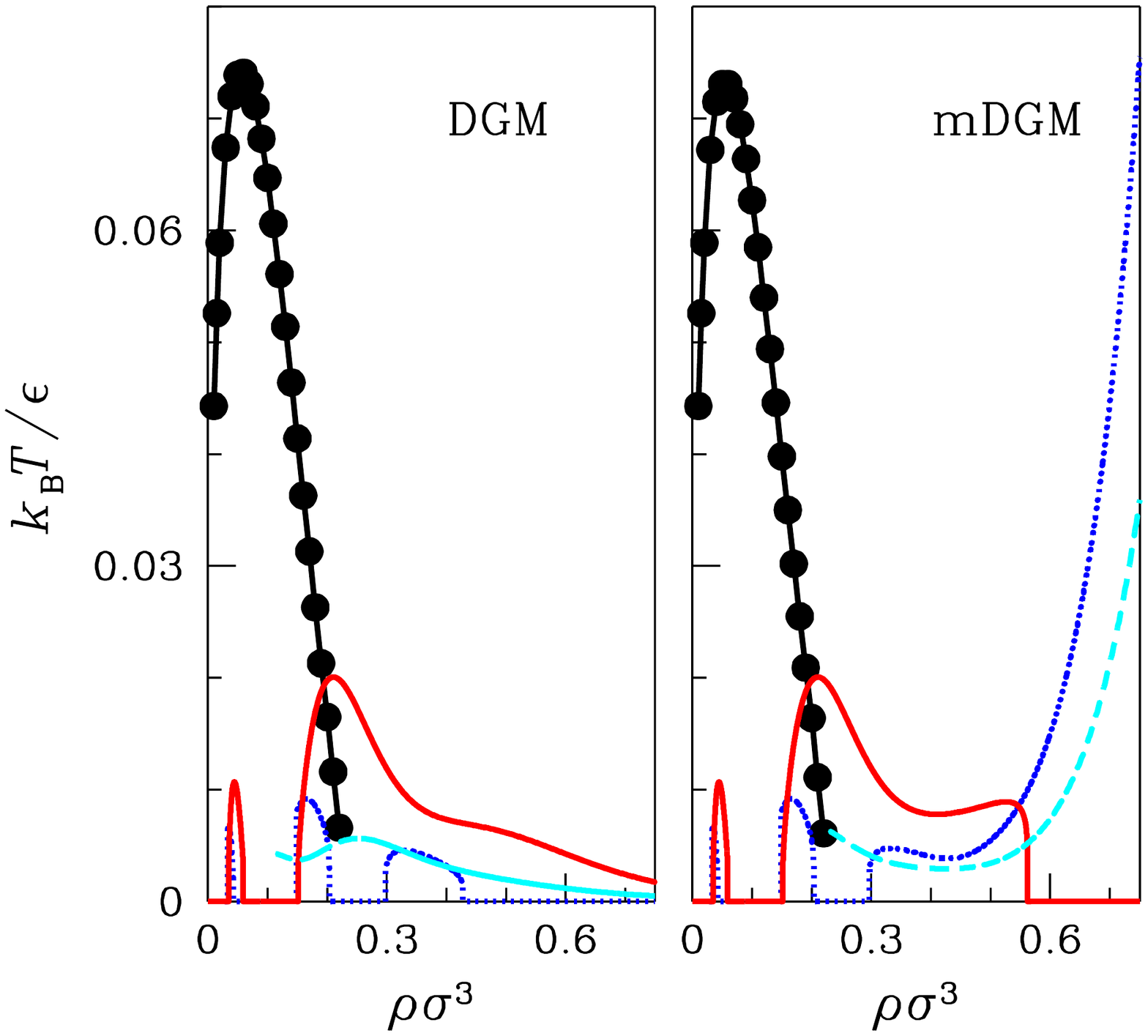}
\caption{
}
\label{fig5}
\end{figure}

\begin{thebibliography}{99}
\bibitem{Likos} C. N. Likos, {\em Phys. Rep.} {\bf 348}, 267 (2001).

\bibitem{Likos4} C. N. Likos, S. Rosenfeldt, N. Dingenouts, M. Ballauff, P. Lindner, N. Werner, and F. V\"{o}gtle, {\em J. Chem. Phys.} {\bf 117}, 1869 (2002).

\bibitem{Bolhuis} P. Bolhuis and D. Frenkel, {\em Phys. Rev. Lett.} {\bf 72}, 2211 (1994).

\bibitem{Franzese} G. Franzese, G. Malescio, A. Skibinsky, S. V. Buldyrev, and H. E. Stanley, {\em Nature} {\em 409}, 692 (2001).

\bibitem{Wilding} N. B. Wilding and J. E. Magee, {\em Phys. Rev. E} {\bf 66}, 031509 (2002).

\bibitem{Gibson} H. M. Gibson and N. B. Wilding, {\em Phys. Rev. E} {\bf 73}, 061507 (2006).

\bibitem{Likos2} C. N. Likos, A. Lang, M. Watzlawek, and H. L\"{o}wen, {\em Phys. Rev. E} {\bf 63}, 031206 (2001).

\bibitem{Prestipino1} S. Prestipino, F. Saija, and P. V. Giaquinta, {\em Phys. Rev. E} {\bf 71}, 050102(R) (2005).

\bibitem{Zhang} K. Zhang, P. Charbonneau, and B. M. Mladek, {\em Phys. Rev. Lett.} {\bf 105}, 245701 (2010).

\bibitem{Louis} A. A. Louis, P. G. Bolhuis, J. P. Hansen, and E. J. Meijer, {\em Phys. Rev. Lett.} {\bf 85}, 2522 (2000).

\bibitem{Likos3} C. N. Likos, {\em Soft Matter} {\bf 2}, 478 (2006).

\bibitem{Fisher} M. E. Fisher and D. Ruelle, {\em J. Math. Phys.} {\bf 7}, 260 (1966).

\bibitem{Ruelle} D. Ruelle, {\em Statistical Mechanics: Rigorous Results} (Imperial College Press, London, 1999).

\bibitem{Heyes} D. M. Heyes and G. Rickayzen, {\em J. Phys.: Condens. Matter} {\bf 19}, 416101 (2007).

\bibitem{Prestipino2} S. Prestipino, C. Speranza, G. Malescio, and P. V. Giaquinta, {\em J. Chem. Phys.} {\bf 140}, 084906 (2014).

\bibitem{Panagiotopoulos} A. Z. Panagiotopoulos, {\em Mol. Phys.} {\bf 61}, 813 (1987).

\bibitem{Widom} B. Widom, {\em J. Chem. Phys.} {\bf 39}, 2802 (1963).

\bibitem{Saija1} F. Saija, S. Prestipino, and G. Malescio, {\em Phys. Rev. E} {\bf 80}, 031502 (2009).

\bibitem{Frenkel} D. Frenkel and B. Smit, {\em Understanding Molecular Simulation}, 2nd ed. (Academic, 2002).

\bibitem{Polson} J. M. Polson, E. Trizac, S. Pronk, and D. Frenkel, {\em J. Chem. Phys.} {\bf 112}, 5339 (2000).

\bibitem{Hansen} See e.g. J.-P. Hansen and I. R. McDonald, {\em Theory of Simple Liquids}, 3rd ed. (Academic, 2006).

\bibitem{Hemmer} P. C. Hemmer and G. Stell, {\em Phys. Rev. Lett.} {\bf 24}, 1284 (1970).

\bibitem{Young} D. A. Young and B. Alder, {\em Phys. Rev. Lett.} {\bf 38}, 1213 (1977).

\bibitem{Debenedetti} P. G. Debenedetti, V. S. Raghavan, and S. S. Borick, {\em J. Phys. Chem.} {\bf 95}, 4540 (1991).

\bibitem{Jagla} E. A. Jagla, {\em Phys. Rev. E} {\bf 63}, 061501 (2001).

\bibitem{Sadr-Lahijani} M. R. Sadr-Lahijany, A. Scala, S. V. Buldyrev, and H. E. Stanley, {\em Phys. Rev. Lett.} {\bf 81}, 4895 (1998).

\bibitem{Mausbach} P. Mausbach and H.-O. May, {\em Fluid Phase Equilibria} {\bf 249}, 17 (2006).

\bibitem{Krekelberg} W. P. Krekelberg, T. Kumar, J. Mittal, J. R. Errington, and T. M. Truskett, {\em Phys. Rev. E} {\bf 79}, 031203 (2009).

\bibitem{Speranza} C. Speranza, S. Prestipino, and P. V. Giaquinta, {\em Mol. Phys.} {\bf 109}, 3001 (2011).

\bibitem{Chaplin} See, for instance, the website entitled ``Water structure and science'' maintained by M. Chaplin at http://www.lsbu.ac.uk/water/index2.html.

\bibitem{Poole} See e.g. P. H. Poole, I. Saika-Voivod, and F. Sciortino, {\em J. Phys.: Condens. Matter} {\bf 17}, L431 (2005).

\bibitem{Wagner} W. Wagner and A. Pruss, {\em J. Phys. Chem. Ref. Data} {\bf 31}, 387 (2002).

\bibitem{Giaquinta1} P. V. Giaquinta and F. Saija, {\em ChemPhysChem} {\bf 6}, 1768 (2005).

\bibitem{Prestipino3} S. Prestipino and P. V. Giaquinta, {\em J. Stat. Phys.} {\bf 96}, 135 (1999).

\bibitem{Prestipino4} S. Prestipino and P. V. Giaquinta, {\em J. Stat. Phys.} {\bf 98}, 507 (2000).

\bibitem{Giaquinta2} P. V. Giaquinta and G. Giunta, {\em Physica A} {\bf 187}, 145 (1992).

\bibitem{Giaquinta3} P. V. Giaquinta, G. Giunta, and G. Malescio, {\em Physica A} {\bf 250}, 91 (1998).
 
\bibitem{Prestipino5} S. Prestipino, {\em J. Phys.: Condens. Matter} {\bf 24}, 035102 (2012).

\bibitem{Prestipino6} S. Prestipino, F. Saija, and G. Malescio, {\em J. Chem. Phys.} {\bf 133}, 144504 (2010).
\end{thebibliography}
\end{document}